\documentclass{PoS}
\usepackage{cite}
\usepackage{amsmath}
\usepackage{amssymb}
\usepackage{epsfig}
\usepackage{etex, xy}
\xyoption{all}

\usepackage{tikz}
\usetikzlibrary{matrix}

\allowdisplaybreaks[4]

\title{
{\rm \footnotesize DESY 17-199, DO-TH 17/33, TTK-17-39, MSUHEP-17-020 
}\\
Heavy Flavor Wilson Coefficients in Deep-Inelastic Scattering: Recent Results\thanks{This work 
was supported in part by the Austrian Science Fund (FWF) grant SFB F50 (F5009-N15), the European 
Commission through contract PITN-GA-2012-316704 ({HIGGSTOOLS}).}
}

\ShortTitle{
Heavy Flavor Wilson Coefficients in Deep-Inelastic Scattering: Recent Results}

\author{J.~Ablinger\\
Research Institute for Symbolic Computation (RISC),
                          Johannes Kepler University, Altenbergerstra\ss{}e 69,
                          A--4040, Linz, Austria}

\author{A.~ Behring, \\
Institut f\"ur Theoretische Teilchenphysik und Kosmologie, RWTH Aachen University,
Sommerfeldstr. 16, D-52074 Aachen, Germany}

\author{
J.~Bl\"umlein\footnote{Speaker}, A.~De Freitas\\
       Deutsches Elektronen-Synchrotron, DESY, Platanenallee 6, D-15738 Zeuthen, Germany}

\author{A.~von Manteuffel\\
Department of Physics and Astronomy, Michigan State University, East Lansing, MI 48824, USA
}

\author{C.~Schneider\\
Research Institute for Symbolic Computation (RISC),
                          Johannes Kepler University, Altenbergerstra\ss{}e 69,
                          A--4040, Linz, Austria}

\abstract{We present recent analytic results for the 3-loop corrections to the massive operator matrix element
$A_{Qg}^{(3)}$ for further color factors. These results have been obtained using the method of arbitrarily large 
moments. We also give an overview on the results which were obtained solving all difference and differential 
equations for the corresponding master integrals that factorize at first order.}

\FullConference{QCDEV2017, JLAB, Newport News, VA, USA, \\ May 22-26, 2017}

\makeatletter
\g@addto@macro\bfseries{\boldmath}
\makeatother
\begin{document}
\section{Introduction}
\label{sec:1}

\vspace*{1mm}
\noindent
{The precision determination of the strong coupling constant $\alpha_s(M_Z^2)$ \cite{ALPHA}, of the parton distribution 
functions \cite{PDF}, and of the heavy quark masses $m_c$ and $m_b$ \cite{MCMB} from the world deep inelastic data 
requires both the knowledge of the NNLO massless and massive QCD corrections to the parton densities. Theses 
quantities are fundamental parameters of the Standard Model or serve as an important input for other measurements 
at hadron colliders to determine further fundamental parameters, like those of the top-quark \cite{TOP} and the Higgs 
sectors \cite{HIGGS}. These are important places at present to find deviations from the Standard Model and therefore
need the most precise possible description.
  
During the last years, essential progress has been made in calculating the massive 3-loop corrections to the 
deep-inelastic structure functions  
\cite{
Blumlein:2006mh,
Bierenbaum:2009mv,
Ablinger:2010ty,
Blumlein:2012vq,
Behring:2014eya,
Ablinger:2014vwa,
Ablinger:2014nga,
Ablinger:2014uka,
Ablinger:2014lka,
Ablinger:2016eyd,
AGG,Behring:2015zaa,Behring:2015roa,
Behring:2016hpa,
Blumlein:2014fqa,
Blumlein:2016xcy}.
Different computational methods  have been worked out in 
Refs.~\cite{Ablinger:2012qm,Ablinger:2014yaa,Ablinger:2015tua}. 
Progress has also been made in the 2-mass case, where several OMEs have been calculated, 
cf.~\cite{Ablinger:2017err,VFNS,APS2,AGG2}. For a recent summary on this see Ref.~\cite{ABILIO}.

All but one of the massive operator matrix elements (OMEs) have been calculated. In this note we describe recent 
developments of the project concerning the calculation of the OME $A_{Qg}^{(3)}$, as well as some technical details of 
the calculation, the present results, and challenges for its completion.

The paper is organized as follows. In Section~\ref{sec:2} we describe the method of arbitrarily 
large moments and calculation techniques for first order factorizing differential and difference 
equations. The contributions to $A_{Qg}^{(3)}$ that have been obtained already by using these 
methods are described in Section~\ref{sec:3}. There we also give an outline on the methods to 
be used to calculate the remaining terms, which are related to new functions, the iterative 
non--iterative integrals \cite{Ablinger:2017bjx}. Section~\ref{sec:4} contains the conclusions.
}
\section{\boldmath The method of arbitrarily large moments and calculation techniques for first order factorizing 
equations}
\label{sec:2}

\vspace*{1mm}
\noindent
In the calculation of the massive OMEs, we reduce the Feynman diagrams to master integrals using integration 
by parts (IBP) relations \cite{IBP}. Depending on their complexity, different calculation methods exist to compute
these integrals. The simplest integrals can be expressed in terms of (generalized) hypergeometric functions
\cite{HYP}. In more general cases, one seeks Mellin-Barnes representations \cite{MB} and uses the residue 
theorem to obtain convergent infinite sum representations. Such sums are also obtained in the case of the 
representation
through hypergeometric functions after expanding in the dimensional parameter $\varepsilon$. 
In Mellin $N$ space, also sum representations are obtained. 
For finite integrals,
one may also use the method of hyperlogarithms \cite{Brown:2008um,Ablinger:2014yaa,FWTHESIS,Panzer:2014gra}.
Having full control on the IBP reduction, one may map divergent integrals into convergent ones 
\cite{vonManteuffel:2014qoa}.
Another powerful tool consists in the method of 
differential equations, cf.~Refs.~\cite{DEQ} and \cite{Ablinger:2015tua}. In the present case, they can be obtained 
from the 
IBP relations by differentiating w.r.t. the resummation parameter $x$ for the operator insertions, 
cf.~\cite{Ablinger:2014lka}. 
Inserting the formal power series in $x$ into the differential equations,
one obtains 
recurrence relations, which, as well as the summation problems mentioned before, 
can be solved by difference field and ring methods 
\cite{Karr:81,Schneider:01,Schneider:05a,
Schneider:07d,Schneider:10b,Schneider:10c,Schneider:15a,Schneider:08c,DFTheory}
using the algorithms 
implemented in the packages {\tt Sigma} \cite{SIG1,SIG2}, {\tt EvaluateMultiSums} and {\tt SumProduction} \cite{EMSSP}.  
In various places the package {\tt HarmonicSums} \cite{Ablinger:PhDThesis,
HARMONICSUMS,
Ablinger:2011te,
Ablinger:2013cf,
Ablinger:2014bra} is used to operate on special functions of different kinds emerging throughout the calculation.
In some cases we also use multi-integration according to the Almkvist-Zeilberger theorem \cite{AZ} 
as an integration method.
This method is implemented in the package {\tt MultiIntegrate} \cite{Ablinger:PhDThesis}. All these methods have been 
described 
in detail in Ref.~\cite{Ablinger:2015tua}. In all cases where difference equations factorize at first order,
{\tt Sigma} will find the solution in Mellin $N$ space. The transition to $x$-space is possible applying algorithms
implemented in the package {\tt HarmonicSums}.

As it is well known, massive single-scale problems starting at 3-loop order are partly related to master integrals
for which the corresponding difference and differential equations do not factorize in first order, 
\cite{Ablinger:2017bjx}.
Yet cases in which this applies to the master integrals, but not to their sum projected on some color-$\zeta$ factors,
i.e. terms which can be distinguished by the corresponding product of Casimir operators and potential (multiple)
zeta values \cite{Blumlein:2009cf}, 
may be dealt with. Considering fixed moments $N$, cf. also \cite{Bierenbaum:2009mv}, the results are given by 
rational numbers and multiple zeta values only \cite{Blumlein:2009cf}. The appearance of new higher functions
like iterated non--iterative integrals manifests only in the corresponding rational sequences but not in new special 
numbers. These sequences can now be analyzed and the difference equations which describe them can be determined
\cite{GUESSHB,Blumlein:2009tj} provided enough Mellin moments can be calculated. This is possible using the algorithm
\cite{Blumlein:2017dxp}, implemented in the package {\tt SolveCoupledSystem}, that can compute
arbitrarily large moments. The corresponding difference 
equations can now be analyzed by the package {\tt Sigma}, which will determine all its 1st order factors, and 
in the present case, a remainder term not factorizing in first order. 
We will illustrate this in the case of the unrenormalized OME
$\hat{\hat{A}}_{Qg}^{(3)}$ in Section~\ref{sec:3}.
\section{\boldmath The contributions to {$\bf A_{Qg}^{(3)}$} through first order factorizing equations}
\label{sec:3}

\vspace*{1mm}
\noindent
{In the following we calculate 3-loop contributions to the massive OME $A_{Qg}^{(3)}$ in the single heavy quark mass 
case. The 1358 Feynman diagrams are generated using {\tt QGRAF} \cite{Nogueira:1991ex} and are brought into a form in 
which the operator insertions \cite{Bierenbaum:2009mv} are resummed into a propagator, cf. \cite{Ablinger:2014lka}, 
using a formal resummation parameter $x$. The color algebra is performed using {\tt Color} 
\cite{vanRitbergen:1998pn}. 
We use {\tt Reduze2} \cite{REDUZE}\footnote{The package uses the codes {\tt GiNaC} \cite{Bauer:2000cp} and {\tt Fermat}
\cite{FERMAT}.} to map the problem to 340 master integrals. 224 master integrals were calculated using the methods 
described in \cite{Ablinger:2015tua}. We derive differential equations for all master integrals, 
which are turned into associated difference equations by the formal power series in $x$ 
\cite{Ablinger:2016yjz}. 
For these difference 
equations we use now the 
technique described in Ref.~\cite{Ablinger:2015tua} and calculate the master integrals for fixed moments $N$, which are 
inserted into the expression of the unrenormalized OME, see Eq.~(4.37) of Ref.~\cite{Bierenbaum:2009mv}. Iteratively, 
we 
calculate higher and higher moments. For the complete expression we have calculated 2000 moments and for the 
contribution with the color factor $O(T_F^2 C_{A,F})$, 8000 moments. The method of guessing 
\cite{GUESSHB,Blumlein:2009tj} 
can now be used to find a difference equation for the different terms according to their contribution in the 
Laurent series in $\varepsilon$ and their color and $\zeta$-value factors.

We first have considered the $O(1/\varepsilon)$ term and extracted the 3-loop anomalous dimension $\gamma_{qg}^{(2)}(N)$ 
in Ref.~\cite{Ablinger:2017tan} solving the difference equations obtained by guessing using the packages 
{\tt Sigma}~\cite{SIG1,SIG2} {\tt EvaluatMultiSum} and {\tt SumProduction} \cite{EMSSP} and by simplifying the 
corresponding expressions using the package {\tt HarmonicSums} 
\cite{Ablinger:PhDThesis,
HARMONICSUMS,
Ablinger:2011te,
Ablinger:2013cf,
Ablinger:2014bra}. 
This is the first independent 
recalculation of this anomalous dimension since it was first computed in \cite{Vogt:2004mw}.

We turn now to the $O(\varepsilon^0)$ terms. In the case of the $O(T_F^2 C_{A,F})$ contributions, the method of 
guessing
has led to difference equations in all cases. The 2000 moments for the remaining color and zeta values were not 
enough, however, to determine difference equations for the pure color factors and those $\propto \zeta_3$, except for 
the $N_F$ terms, for which the number of moments had been sufficient and which also had been calculated before in 
Ref.~\cite{Ablinger:2010ty}. Analyzing the terms with pure color factors and those $\propto \zeta_3$ for the 
contributions 
$O(T_F^2 C_{A,F})$ using {\tt Sigma}, it turned out that the corresponding difference equations do not factorize in 
first 
order completely. Already here we speak of difference equations of order $o = 45$ and degree $d \sim 1500$.
They could be reduced to low order non-factorizing difference equations. The reason for this is that, 
as we know through 
other analyses, elliptic parts are contained in these cases and one expects iterative non-iterative integral 
solutions here, cf.~\cite{Ablinger:2017bjx}.

For quite a series of color and $\zeta$-factors, namely all but 10, the constant part of the unrenormalized OME
$\hat{\hat{A}}_{Qg}^{(3)}$, $a_{Qg}^{(3)}(N)$, could be calculated using the above methods. It is given by 
\begin{eqnarray}
a_{Qg}^{(3)}(N) &=& \textcolor{blue}{C_A^2 T_F} \Biggl\{ t_{C_A^2 T_F}(N) + t_{C_A^2 T_F \zeta_3}(N) \zeta_3 
\nonumber\\ && 
+ \frac{72 (-2+N) (3+N) P_2}{(N-1) N^2 (1+N)^2 (2+N)^2} \zeta_4
        -\frac{4 P_{17}}{(N-1) N^2 (1+N)^2 (2+N)^2} {\sf B_4}
\nonumber\\ &&
        +p_{qg}^{(0)} \Biggl[
                -16 {\sf B_4} S_1
                +144 \zeta_4 S_1
                +\Biggl(
                        -16 S_1^3
                        -\frac{4 P_7}{3 (N-1) N (1+N) (2+N)} S_2
                        -32 S_1 S_2
\nonumber\\ &&
                        -8 S_3
                        +\Biggl(
                                -\frac{8 P_8}{3 (N-1) N (1+N) (2+N)}
                                -48 S_1
                        \Biggr) S_{-2}
                        -8 S_{-3}
                        +16 S_{-2,1}
                \Biggr) \zeta_2
        \Biggr]
\nonumber\\ &&
        +\Biggl[
                \frac{2 P_{35}}{9 (N-1)^2 N^4 (1+N)^4 (2+N)^4}
                +\frac{4 P_{31}}{9 (N-1)^2 N^3 (1+N)^3 (2+N)^3} S_1
\nonumber\\ &&
                -\frac{4 P_{18}}{3 (N-1) N^2 (1+N)^2 (2+N)^2} S_1^2
        \Biggr] \zeta_2
\Biggr\}
+ \textcolor{blue}{C_A T_F^2} \Biggl\{ t_{C_A T_F^2}(N) + t_{C_A T_F^2 \zeta_3}(N) \zeta_3
\nonumber\\ &&
+ \textcolor{blue}{N_F}
        \Biggl\{
                -\frac{8 P_{36}}{243 (N-1) N^5 (1+N)^5 (2+N)^5}
                + p_{qg}^{(0)} \Biggl[
                        \Biggl(
                                \frac{1888}{27} S_3
                                +\frac{224}{9} S_{2,1}
                        \Biggr) S_1
\nonumber\\ &&
                        +\frac{32}{27} S_1^4
                        +\frac{176}{9} S_1^2 S_2
                        +\frac{80}{9} S_2^2
                        +\frac{640}{9} S_4
                        +\Biggl(
                                -\frac{64 (2 N-1)}{(N-1) N} S_1
                                +\frac{128}{3} S_2
                        \Biggr) S_{-2}
                        +\frac{64}{9} S_{-4}
\nonumber\\ &&
                        -\frac{32}{3} S_{1,-3}
                        -
                        \frac{64}{3} S_{2,-2}
                        -\frac{32}{9} S_{3,1}
                        +\frac{64 (2 N-1)}{(N-1) N} S_{-2,1}
                        +64 S_{1,1,-2}
                        -\frac{416}{9} S_{2,1,1}
\nonumber\\ &&
                        +\Biggl(
                                \frac{16}{3} S_1^2
                                +\frac{16}{3} S_2
                                +\frac{32}{3} S_{-2}
                        \Biggr) \zeta_2
                        +\Biggl(
                                \frac{448 \big(
                                        1+N+N^2\big)}{9 (N-1) N (1+N) (2+N)}
                                -\frac{224}{9} S_1
                        \Biggr) \zeta_3
                \Biggr]
\nonumber\\ &&
                +\Biggl(
                        \frac{16 P_{32}}{243 (N-1) N^2 (1+N)^4 (2+N)^4}
                        -\frac{16 P_{10}}{27 N (1+N)^2 (2+N)^2} S_2
                \Biggr) S_1
\nonumber\\ &&
                +\frac{8 P_{20}}{81 N (1+N)^3 (2+N)^3} S_1^2
                -\frac{16 P_9}{81 N (1+N)^2 (2+N)^2} S_1^3
\nonumber\\ &&
                +\frac{8 P_{26}}{81 (N-1) N^3 (1+N)^3 (2+N)^3} S_2
                -\frac{32 P_{22}}{81 (N-1) N^2 (1+N)^2 (2+N)^2} S_3
\nonumber\\ &&
                +\frac{32 P_{21}}{81 N (1+N)^3 (2+N)^3} S_{-2}
                +\frac{32 P_{15}}{27 (N-1) N^2 (1+N)^2 (2+N)^2} S_{-3}
\nonumber\\ &&
                -\frac{64 P_{16}}{9 (N-1) N^2 (1+N)^2 (2+N)^2} S_{1,-2}
                -\frac{64 P_4}{27 N (1+N)^2 (2+N)^2} S_{2,1}
\nonumber\\ &&
                +\Biggl(
                        -\frac{4 P_{27}}{9 (N-1) N^3 (1+N)^3 (2+N)^3}
                        -\frac{16 P_5}{9 N (1+N)^2 (2+N)^2} S_1
                \Biggr) \zeta_2
        \Biggr\} 
\nonumber\\ &&
        +\Biggl(
                -\frac{4 P_{28}
                }{9 (N-1) N^3 (1+N)^3 (2+N)^3}
                +
                \frac{160 \big(
                        4-N+N^2+4 N^3+N^4\big)}{9 N (1+N)^2 (2+N)^2} S_1
        \Biggr) \zeta_2
\nonumber\\ &&
        + p_{qg}^{(0)} \Biggl(
                \frac{40}{3} S_1^2
                +\frac{40}{3} S_2
                +\frac{80}{3} S_{-2}
        \Biggr) \zeta_2
\Biggr\}
+ \textcolor{blue}{C_F^2 T_F}  \Biggl\{ t_{C_F^2 T_F}(N) + t_{C_F^2 T_F \zeta_3}(N) \zeta_3
\nonumber\\ &&
        -\frac{16 (N-1) \big(
                -2+3 N+3 N^2\big)}{N^2 (1+N)^2} {\sf B_4}
        +\frac{72 (N-1) \big(
                -2+3 N+3 N^2\big)}{N^2 (1+N)^2} \zeta_4
\nonumber\\ && 
        +\Biggl[
                \frac{P_{23}}{2 N^4 (1+N)^4 (2+N)}
                +\frac{8 P_{19}}{N^3 (1+N)^3 (2+N)} S_1
                +\frac{4 P_3}{N^2 (1+N)^2 (2+N)} S_1^2
        \Biggr] \zeta_2
\nonumber\\ && 
        + p_{qg}^{(0)} \Biggl[
                -16 S_1^3
                -\frac{8 \big(
                        2+3 N+3 N^2\big)}{N (1+N)} S_2
                +32 S_1 S_2
                +16 S_3
\nonumber\\ && 
                +\Biggl(
                        -\frac{16}{N (1+N)}
                        +32 S_1
                \Biggr) S_{-2}
                +16 S_{-3}
                -32 S_{-2,1}
        \Biggr] \zeta_2
\Biggr\}
+ \textcolor{blue}{C_F} \Biggl\{ \textcolor{blue}{C_A T_F} \Biggl\{ t_{C_F C_A T_F}(N) + 
\nonumber\\ &&
t_{C_F C_A T_F \zeta_3}(N) \zeta_3 
                + \frac{32 P_{13}}{(N-1) N^2 (1+N)^2 (2+N)^2} \left({\sf B_4} - \frac{9}{2} \zeta_4\right)
\nonumber\\ && 
                + p_{qg}^{(0)} \Biggl[
                        32 {\sf B_4} S_1 
                        -144 \zeta_4 S_1
                        +\Biggl(
                                32 S_1^3
                                -\frac{12 P_1}{(N-1) N (1+N) (2+N)} S_2
                                -8 S_3
\nonumber\\ && 
                                +\Biggl(
                                        -
                                        \frac{8 \big(
                                                1+3 N+3 N^2\big)}{N (1+N)}
                                        +16 S_1
                                \Biggr) S_{-2}
                                -8 S_{-3}
                                +16 S_{-2,1}
                        \Biggr) \zeta_2 
                \Biggr] 
\nonumber\\ && 
                +\Biggl(
                        \frac{P_{25}}{18 (N-1) N^3 (1+N)^3 (2+N)^3}
                        -\frac{4 P_{29}}{9 (N-1) N^3 (1+N)^3 (2+N)^3} S_1
\nonumber\\ &&
                        +\frac{8 P_{12}}{3 (N-1) N^2 (1+N)^2 (2+N)^2} S_1^2
                \Biggr) \zeta_2
        \Biggr\}
+ \textcolor{blue}{T_F^2} \Biggl\{ t_{C_F T_F^2}(N) + t_{C_F T_F^2 \zeta_3}(N) \zeta_3
\nonumber\\ &&
              + \textcolor{blue}{N_F} \Biggl\{
                        \frac{P_{37}}{243 (-1+N) N^6 (1+N)^6 (2+N)^5}
                        + p_{qg}^{(0)} \Biggl[
                                \Biggl(
                                        -\frac{256}{27} S_3
                                        -\frac{128}{3} S_{2,1}
                                \Biggr) S_1
\nonumber\\ && 
                                -\frac{32}{27} S_1^4
                                -\frac{64}{9} S_1^2 S_2
                                -\frac{128}{9} S_2^2
                                +\frac{256}{9} S_4
                                -\frac{128}{3} S_{3,1}
                                +\frac{256}{3} S_{2,1,1}
                                -\frac{16}{3} S_1^2 \zeta_2
\nonumber\\ && 
                                +\Biggl(
                                        -\frac{56 P_{14}}{9 (-1+N) N^2 (1+N)^2 (2+N)}
                                        +\frac{224}{9} S_1
                                \Biggr) \zeta_3
                        \Biggr]
                        +\Biggl(
                                -\frac{16 P_{11}}{243 N^2 (1+N)^3 (2+N)}
\nonumber\\ && 
                                +\frac{32 \big(
                                        24+83 N+49 N^2+10 N^3\big)}{27 N^2 (1+N) (2+N)} S_2
                        \Biggr) S_1
                        -\frac{32 P_6}{81 N^2 (1+N)^2 (2+N)} S_1^2
\nonumber\\ && 
                        +\frac{32 \big(
                                24+83 N+49 N^2+10 N^3\big)}{81 N^2 (1+N) (2+N)}
                         S_1^3
                        +\frac{8 P_{33}}{27 (-1+N) N^4 (1+N)^4 (2+N)^3} S_2
\nonumber\\ && 
                        -\frac{16 P_{24}}{81 (-1+N) N^3 (1+N)^3 (2+N)^2} S_3
                        -\frac{128 \big(
                                -2-3 N+N^2\big)}{3 N^2 (1+N) (2+N)} S_{2,1}
\nonumber\\ &&
                        +\Biggl(
                                \frac{2 (-2+N) P_{30}}{9 (-1+N) N^4 (1+N)^4 (2+N)^3}
                                +\frac{16 \big(
                                        12+28 N+11 N^2+5 N^3\big)}{9 N^2 (1+N) (2+N)} S_1
                        \Biggr) \zeta_2
                \Biggr\} 
\nonumber\\ && 
                +\Biggl(
                        \frac{2 P_{34}}{9 (-1+N) N^4 (1+N)^4 (2+N)^3}
                        +\frac{80 \big(
                                6+11 N+4 N^2+N^3\big)}{9 N^2 (1+N) (2+N)} S_1
                \Biggr) \zeta_2
\nonumber\\ &&
                + p_{qg}^{(0)} \Biggl(
                        -\frac{40}{3} S_1^2
                        +8 S_2
                \Biggr) \zeta_2
        \Biggr\}
\Biggr\}
-\frac{64}{9} p_{qg}^{(0)} \textcolor{blue}{T_F^3} \zeta_3.
\end{eqnarray}
Here we used the shorthand notation
\begin{eqnarray}
p_{qg}^{(0)}(N) = \frac{N^2+N+2}{(N + 2) (N + 1) N},
\end{eqnarray}
$S_{\vec{a}}$ denote the nested harmonic sums \cite{HSUM}
\begin{eqnarray}
S_{b,\vec{a}} \equiv S_{b,\vec{a}}(N) = \sum_{k=1}^N \frac{({\rm sign}(b))^k}{k^{|b|}} S_{\vec{a}}(k),
~~~S_\emptyset = 1,~~b, a_i \in \mathbb{Z} \backslash \{0\},
\end{eqnarray}
the constant ${\sf B_4}$ is
\begin{eqnarray}
{\sf B_4} = - 4 \zeta_2 \ln^2(2) + \frac{2}{3} \ln^4(2) - \frac{13}{2} \zeta_4 + 16 \text{Li}_4\left(\frac{1}{2}\right),
\end{eqnarray}
and $P_i$ denote polynomials in $N$ (which have be computed explicitly), and the functions $t_j(N)$ have still to 
be calculated. Using the methods 
of Ref.~\cite{Ablinger:2015tua}, we have calculated a lot more Feynman diagrams for the color factors 
contributing to $\hat{\hat{A}}_{Qg}^{(3)}$, which could not yet been gotten using the method of arbitrarily large 
moments, so that 1122 of 1358 diagrams have been calculated by now.
}
\section{Conclusions}
\label{sec:4}

\vspace*{1mm}
\noindent
Advanced methods in calculating Feynman and master integrals for massive 3-loop OMEs 
allowed us already to compute a significant part of 
the contributions to $a_{Qg}^{(3)}$ analytically. Our toolbox employs various methods \cite{Ablinger:2015tua}, and 
relies in particular on very
efficient solvers based on difference field and ring theory 
\cite{Karr:81,Schneider:01,Schneider:05a,Schneider:07d,Schneider:10b,Schneider:10c,
Schneider:15a,Schneider:08c,DFTheory}.  These algorithms have been implemented
in the packages {\tt Sigma} \cite{SIG1,SIG2}, {\tt EvaluateMultiSums} and {\tt SumProduction} \cite{EMSSP}.
A very efficient treatment of special functions has been possible using the package
{\tt HarmonicSums} \cite{Ablinger:PhDThesis,
HARMONICSUMS,
Ablinger:2011te,
Ablinger:2013cf,
Ablinger:2014bra}.
The method of arbitrarily large moments \cite{Blumlein:2017dxp} could now be used to 
calculate the 3-loop anomalous dimension $\gamma_{qg}^{(2)}(N)$ automatically from first principles even in the more 
involved massive environment \cite{Ablinger:2017tan}. The same method allowed to calculate  the contributions to the 
$O(\varepsilon^0$) term $a_{Qg}^{(3)}$ for 18 out of 28 of the color-$\zeta$ terms in analytic form. 

Including the use of other methods described in \cite{Ablinger:2015tua}, 1122 of 1358 Feynman diagrams contributing 
to $A_{Qg}^{(3)}$ at $O(\varepsilon^0)$ have been calculated. The remaining terms contain iterative non-iterative 
integrals over also elliptic letters and are currently being computed.



\begin{thebibliography}{99}
%
\bibitem{ALPHA}
  S.~Bethke {\it  et al.}, 
  {\it  Workshop on Precision Measurements of $\alpha_s$},
  arXiv:1110.0016 [hep-ph];\\
  S.~Moch, S.~Weinzierl {\it  et al.}, 
  {\it  High precision fundamental constants at the TeV scale},
  arXiv:1405.4781 [hep-ph];\\
  S.~Alekhin, J.~Bl\"umlein and S.O.~Moch,
  Mod.\ Phys.\ Lett.\ A {\bf 31} (2016) no.25,  1630023.
%
\bibitem{PDF}
  A.~Accardi {\it  et al.},
  Eur.\ Phys.\ J.\ C {\bf 76} (2016) no.8,  471
  [arXiv:1603.08906 [hep-ph]];\\
  S.~Alekhin, J.~Bl\"umlein, S.~Moch and R.~Placakyte,
  Phys.\ Rev.\ D {\bf 96} (2017) no.1,  014011
  [arXiv:1701.05838 [hep-ph]].
%
\bibitem{MCMB}
  S.~Alekhin, J.~Bl\"umlein, K.~Daum, K.~Lipka and S.~Moch,
  Phys.\ Lett.\ B {\bf 720} (2013) 172
  [arXiv:1212.2355 [hep-ph]];\\
  A.~Gizhko {\it et al.},
  Phys.\ Lett.\ B {\bf 775} (2017) 233
  [arXiv:1705.08863 [hep-ph]].
%
\bibitem{TOP}
  M.~Czakon, P.~Fiedler and A.~Mitov,
  Phys.\ Rev.\ Lett.\  {\bf 110} (2013) 252004
  [arXiv:1303.6254 [hep-ph]].
%
\bibitem{HIGGS}
  C.~Anastasiou, C.~Duhr, F.~Dulat, E.~Furlan, T.~Gehrmann, F.~Herzog, A.~Lazopoulos and B.~Mistlberger,
  JHEP {\bf 1605} (2016) 058
  [arXiv:1602.00695 [hep-ph]].
%
\bibitem{Blumlein:2006mh}
  J.~Bl\"umlein, A.~De Freitas, W.L.~van Neerven and S.~Klein,
  Nucl.\ Phys.\ B {\bf 755} (2006) 272
  [hep-ph/0608024].
%
\bibitem{Bierenbaum:2009mv}
  I.~Bierenbaum, J.~Bl\"umlein and S.~Klein,
  Nucl.\ Phys.\ B {\bf 820} (2009) 417
  [arXiv:0904.3563 [hep-ph]];\\
  J.~Bl\"umlein, S.~Klein and B.~T\"odtli,
  Phys.\ Rev.\ D {\bf 80} (2009) 094010
  [arXiv:0909.1547 [hep-ph]].
%
\bibitem{Ablinger:2010ty}
  J.~Ablinger, J.~Bl\"umlein, S.~Klein, C.~Schneider and F.~Wi{\ss}brock,
  Nucl.\ Phys.\ B {\bf 844} (2011) 26
  [arXiv:1008.3347 [hep-ph]].
%
\bibitem{Blumlein:2012vq}
  J.~Bl\"umlein, A.~Hasselhuhn, S.~Klein and C.~Schneider,
  Nucl.\ Phys.\ B {\bf 866} (2013) 196
  [arXiv:1205.4184 [hep-ph]].
%
\bibitem{Behring:2014eya}
  A.~Behring, I.~Bierenbaum, J.~Bl\"umlein, A.~De Freitas, S.~Klein and F.~Wi{\ss}brock,
  Eur.\ Phys.\ J.\ C {\bf 74} (2014) no.9,  3033
  [arXiv:1403.6356 [hep-ph]].
%
\bibitem{Ablinger:2014vwa}
J.~Ablinger, A.~Behring, J.~Bl\"umlein, A.~De Freitas, A. Hasselhuhn, A.~von Manteuffel, M.~Round, 
C.~Schneider, and F.~Wi{\ss}brock,
  Nucl.\ Phys.\ B {\bf 886} (2014) 733
  [arXiv:1406.4654 [hep-ph]].
%
\bibitem{Ablinger:2014nga}
  J.~Ablinger, A.~Behring, J.~Bl\"umlein, A.~De Freitas, A.~von Manteuffel and C.~Schneider,
  Nucl.\ Phys.\ B {\bf 890} (2014) 48
  [arXiv:1409.1135 [hep-ph]].
%
\bibitem{Ablinger:2014uka}
  J.~Ablinger, J.~Bl\"umlein, A.~De Freitas, A.~Hasselhuhn, A.~von Manteuffel, M.~Round and C.~Schneider,
  Nucl.\ Phys.\ B {\bf 885} (2014) 280
  [arXiv:1405.4259 [hep-ph]].
%
\bibitem{Ablinger:2014lka}
  J.~Ablinger, J.~Bl\"umlein, A.~De Freitas, A.~Hasselhuhn, A.~von Manteuffel, M.~Round, C.~Schneider and F.~Wi{\ss}brock,
  Nucl.\ Phys.\ B {\bf 882} (2014) 263
  [arXiv:1402.0359 [hep-ph]].
%
\bibitem{Ablinger:2016eyd}
  J.~Ablinger, A.~Behring, J.~Bl\"umlein, A.~De Freitas, A.~Hasselhuhn, A.~von~Manteuffel, C.G.~Raab, M.~Round, 
  C.~Schneider and F.~Wi\ss{}brock, 
  PoS (EPS-HEP2015) (2015) 504
  [arXiv:1602.00583 [hep-ph]]. 	
%
\bibitem{AGG}
  J.~Ablinger et al., DESY 15--112.
%
\bibitem{Behring:2015zaa}
  A.~Behring, J.~Bl\"umlein, A.~De Freitas, A.~von Manteuffel and C.~Schneider,
  Nucl.\ Phys.\ B {\bf 897} (2015) 612
  [arXiv:1504.08217 [hep-ph]].
%
\bibitem{Behring:2015roa}
  A.~Behring, J.~Bl\"umlein, A.~De Freitas, A.~Hasselhuhn, A.~von Manteuffel and C.~Schneider,
  Phys.\ Rev.\ D {\bf 92} (2015) no.11,  114005
  [arXiv:1508.01449 [hep-ph]].
%
\bibitem{Behring:2016hpa}
  A.~Behring, J.~Bl\"umlein, G.~Falcioni, A.~De Freitas, A.~von Manteuffel and C.~Schneider,
  Phys.\ Rev.\ D {\bf 94} (2016) no.11,  114006
  [arXiv:1609.06255 [hep-ph]].
%
\bibitem{Blumlein:2014fqa}
  J.~Bl\"umlein, A.~Hasselhuhn and T.~Pfoh,
  Nucl.\ Phys.\ B {\bf 881} (2014) 1
  [arXiv:1401.4352 [hep-ph]].
%
\bibitem{Blumlein:2016xcy}
  J.~Bl\"umlein, G.~Falcioni and A.~De Freitas,
  Nucl.\ Phys.\ B {\bf 910} (2016) 568
  [arXiv:1605.05541 [hep-ph]].
%
\bibitem{Ablinger:2012qm}
  J.~Ablinger, J.~Bl\"umlein, A.~Hasselhuhn, S.~Klein, C.~Schneider and F.~Wi{\ss}brock,
  Nucl.\ Phys.\ B {\bf 864} (2012) 52
  [arXiv:1206.2252 [hep-ph]].
%
\bibitem{Ablinger:2014yaa}
  J.~Ablinger, J.~Bl\"umlein, C.~Raab, C.~Schneider and F.~Wi\ss{}brock,
  Nucl.\ Phys.\ B {\bf 885} (2014) 409
  [arXiv:1403.1137 [hep-ph]].
%
\bibitem{Ablinger:2015tua}
  J.~Ablinger, A.~Behring, J.~Bl\"umlein, A.~De Freitas, A.~von Manteuffel and C.~Schneider,
  Comput.\ Phys.\ Commun.\  {\bf 202} (2016) 33
  [arXiv:1509.08324 [hep-ph]].
%
\bibitem{Ablinger:2017err}
  J.~Ablinger, J.~Bl\"umlein, A.~De Freitas, A.~Hasselhuhn, C.~Schneider and F.~Wi{\ss}brock,
  Nucl.\ Phys.\ B {\bf 921} (2017) 585
  [arXiv:1705.07030[hep-ph]].
%
\bibitem{APS2}
  J.~Ablinger, J.~Bl\"umlein, A.~De Freitas, C.~Schneider and K.~Sch\"onwald,
  {\it The two-mass contribution to the three-loop pure singlet operator matrix element},
  arXiv:1711.06717 [hep-ph].
%
\bibitem{VFNS}
J.~Bl\"umlein, A.~De Freitas, C.~Schneider, and K.~Sch\"onwald, DESY 17-187.
%
\bibitem{AGG2}
J.~Ablinger, J.~Bl\"umlein, A.~De Freitas, A.~Goedicke, C.~Schneider and K.~Sch\"onwald, in preparation.
%
\bibitem{ABILIO}
J.~Ablinger et al., DO-TH 17/34.
%
\bibitem{Ablinger:2017bjx}
  J.~Ablinger, J.~Bl\"umlein, A.~De Freitas, M.~van Hoeij, E.~Imamoglu, C.~G.~Raab, C.-S.~Radu and C.~Schneider,
  arXiv:1706.01299 [hep-th].
%
\bibitem{IBP}
J. Lagrange, {\it Nouvelles recherches sur la nature et la propagation
du son}, Miscellanea Taurinensis, t. II, 1760-61; Oeuvres t. I, p. 263;\\
C.F. Gau\ss{}, {Theoria attractionis corporum sphaeroidicorum ellipticorum
homogeneorum methodo novo tractate}, Commentationes societas scientiarum
Gottingensis recentiores, Vol III, 1813, Werke Bd. {\bf V} pp. 5-7;\\
G. Green, {\it Essay on the Mathematical Theory of Electricity and
Magnetism}, Nottingham, 1828 [Green Papers, pp. 1-115];\\
M. Ostrogradski, Mem. Ac. Sci. St. Peters., {\bf 6}, (1831) 39;\\
  K.G.~Chetyrkin and F.V.~Tkachov,
  Nucl.\ Phys.\ B {\bf 192} (1981) 159.
%
\bibitem{HYP}
W.N.~Bailey, {\it Generalized Hypergeometric Series}, (Cambridge University
Press,  Cambridge, 1935);\\
L.J.~Slater, {\it Generalized Hypergeometric Functions}, (Cambridge University
Press, Cambridge, 1966);\\
P.~Appell and J.~Kamp\'{e} de F\'{e}riet, {\it Fonctions
Hyperg\'{e}om\'{e}triques et Hypersph\'{e}riques, Polynomes d' Hermite},
(Gauthier-Villars, Paris, 1926);\\
P.~Appell, {\it Les Fonctions Hyperg\'{e}om\'{e}triques de Plusieur
Variables}, (Gauthier-Villars, Paris, 1925);\\
J.~Kamp\'{e} de F\'{e}riet, {\it La fonction
hyperg\'{e}om\'{e}trique}, (Gauthier-Villars, Paris, 1937);\\
H. Exton, {\it Multiple Hypergeometric Functions and Applications},
(Ellis Horwood, Chichester, 1976);\\
H.~Exton, {\it Handbook of Hypergeometric Integrals},
(Ellis Horwood, Chichester, 1978);\\
H.M.~Srivastava and P.W.~Karlsson, {\it Multiple Gaussian Hypergeometric
Series}, (Ellis Horwood, Chicester, 1985);\\
  M.J.~Schlosser, in: {\it Computer Algebra in Quantum Field Theory: Integration, Summation and
  Special Functions}, C. Schneider, J. Bl\"umlein, Eds.,~p.~305, (Springer, Wien, 2013)
  [arXiv:1305.1966 [math.CA]].
%
\bibitem{MB}
E.W.~Barnes, 
Proc. Lond. Math. Soc. (2) {\bf 6} (1908) 141; 
Quart.
Journ. Math. {\bf 41} (1910) 136;\\ 
H.~Mellin,
Math. Ann. {\bf 68} (1910) 305;\\
  M.~Czakon,
  Comput.\ Phys.\ Commun.\  {\bf 175} (2006) 559
  [hep-ph/0511200];\\
  A.V.~Smirnov and V.A.~Smirnov,
  Eur.\ Phys.\ J.\ C {\bf 62} (2009) 445
  [arXiv:0901.0386 [hep-ph]].
%
\bibitem{Brown:2008um}
  F.C.S.~Brown, 
  Commun.\ Math.\ Phys.\  {\bf 287} (2009) 925  
  [arXiv:0804.1660 [math.AG]].
%
\bibitem{FWTHESIS}
F.~Wi\ss{}brock, 
{\it $O(\alpha_s^3)$ Contributions to the Heavy Flavor Wilson Coefficients of the Structure Function $F_2(x,Q^2)$
at $Q^2 \gg m^2$}, 
PhD Thesis, TU Dortmund, 2015.
%
\bibitem{Panzer:2014gra}
  E.~Panzer,
  JHEP {\bf 1403} (2014) 071
  [arXiv:1401.4361 [hep-th]].
%
\bibitem{vonManteuffel:2014qoa}
  A.~von Manteuffel, E.~Panzer and R.M.~Schabinger,
  JHEP {\bf 1502} (2015) 120
  [arXiv:1411.7392 [hep-ph]].
%
\bibitem{DEQ}
  A.V.~Kotikov,
  Phys.\ Lett.\ B {\bf 254} (1991) 158;\\
  E.~Remiddi,
  Nuovo Cim.\ A {\bf 110} (1997) 1435
  [hep-th/9711188];\\
  M.~Caffo, H.~Czyz, S.~Laporta and E.~Remiddi,
  Acta Phys.\ Polon.\ B {\bf 29} (1998) 2627
  [hep-th/9807119];
  Nuovo Cim.\ A {\bf 111} (1998) 365
  [hep-th/9805118];\\
  T.~Gehrmann and E.~Remiddi,
  Nucl.\ Phys.\ B {\bf 580} (2000) 485
  [hep-ph/9912329].
%
\bibitem{Karr:81}
M.~Karr, 
{J.~ACM} {\bf 28} (1981) 305.
%
\bibitem{Schneider:01}
C.~Schneider,
RISC, Johannes Kepler University, Linz technical report 01-17 (2001).
%
\bibitem{Schneider:05a}
C. Schneider,
An. Univ. Timisoara Ser. Mat.-Inform. {\bf 42} (2004) 163;\\
{J. Differ. Equations Appl.\/} {\bf 11} (2005) 799
;\\
Appl. Algebra Engrg. Comm. Comput. {\bf 16} (2005) 1.
%
\bibitem{Schneider:07d}
C.~Schneider, 
{J. Algebra Appl.\/} {\bf 6} (2007) 415.
%
\bibitem{Schneider:10b}
C.~Schneider, {\it {Motives, Quantum Field Theory, and Pseudodifferential
  Operators}\/} ({\it Clay Mathematics Proceedings\/} Vol.~{\bf{12}} ed. A.~Carey,
  D.~Ellwood, S.~Paycha and S.~Rosenberg,(Amer. Math. Soc) (2010), 285 
  [arXiv:0904.2323].
%
\bibitem{Schneider:10c}
C.~Schneider, 
{Ann. Comb.\/} {\bf 14} (2010) 533 
[arXiv:0808.2596].
%
\bibitem{Schneider:15a}
C.~Schneider, 
in: Computer Algebra and Polynomials, Applications of Algebra and Number Theory, J.~Gutierrez, J.~Schicho, M.~Weimann (ed.), Lecture Notes in Computer Science (LNCS) 8942 (2015), 157
[arXiv:13077887 [cs.SC]].

\bibitem{Schneider:08c}
C. Schneider, 
J. Symbolic Comput. {\bf 43} (2008) 611 
[arXiv:0808.2543v1];
J. Symb. Comput. {\bf 72} (2016) 82
[arXiv:1408.2776 [cs.SC]];
J. Symb. Comput. 80 (2017), 616  [arXiv:1603.04285 [cs.SC]].
%
\bibitem{DFTheory}
C.~Schneider, 
Ann. Comb. \textbf{9}(1) (2005) 75; 
\\
S.A. Abramov and M.~Petkov{\v{s}}ek,
{J. Symbolic Comput.}, {\bf 45}(6) (2010) 684; 
\\
C.~Schneider,
{\it Structural theorems for symbolic summation},
{Appl. Algebra Engrg. Comm. Comput.}, {\bf 21}(1) (2010) 1;
\\
C.~Schneider, 
In: Symbolic and Numeric Algorithms for Scientific Computing (SYNASC), 2014, 15th International 
Symposium, F.~Winkler, V.~Negru, T.~Ida, T.~Jebelean, D.~Petcu, S.~Watt, D.~Zaharie (ed.), 
(2015)~pp.~26;
IEEE Computer Society, arXiv:1412.2782v1 [cs.SC].
%
\bibitem{SIG1}
C.~Schneider, {S\'em.~Lothar. Combin.\/} {\bf 56} (2007) 1, 
 article B56b.
%
\bibitem{SIG2}
C.~Schneider, {{\it Computer Algebra in Quantum Field Theory: Integration,
  Summation and Special Functions}\/} Texts and Monographs in Symbolic
  Computation eds. C.~Schneider and J.~Bl\"umlein  (Springer, Wien, 2013) 325, 
  arXiv:1304.4134 [cs.SC].
%
\bibitem{EMSSP}
  J.~Ablinger, J.~Bl\"umlein, S.~Klein and C.~Schneider,
  Nucl.\ Phys.\ Proc.\ Suppl.\  {\bf 205-206} (2010) 110
  [arXiv:1006.4797 [math-ph]];\\
  J.~Bl\"umlein, A.~Hasselhuhn and C.~Schneider,
  PoS (RADCOR 2011) 032
  [arXiv:1202.4303 [math-ph]];\\
  C.~Schneider,
  J.\ Phys.\ Conf.\ Ser.\  {\bf 523} (2014) 012037
  [arXiv:1310.0160 [cs.SC]].
%
\bibitem{Ablinger:PhDThesis}
  J.~Ablinger,
  {\it Computer Algebra Algorithms for Special Functions in Particle Physics},
  Ph.D. Thesis, J. Kepler University Linz, 2012,
  arXiv:1305.0687 [math-ph].
%
\bibitem{HARMONICSUMS}
  J.~Ablinger,
  PoS {(LL2014)} 019;
  {\it Computer Algebra Algorithms for Special Functions in Particle Physics}, Ph.D. Thesis, J. Kepler University 
Linz, 2012,
  arXiv:1305.0687 [math-ph];\\
  {\it A Computer Algebra Toolbox for Harmonic Sums Related to Particle Physics}, Diploma Thesis, J. Kepler University 
Linz, 2009,
  arXiv:1011.1176 [math-ph].
%
\bibitem{Ablinger:2011te}
  J.~Ablinger, J.~Bl\"umlein and C.~Schneider,
  J.\ Math.\ Phys.\  {\bf 52} (2011) 102301
  [arXiv:1105.6063 [math-ph]].
%
\bibitem{Ablinger:2013cf}
  J.~Ablinger, J.~Bl\"umlein and C.~Schneider,
  J.\ Math.\ Phys.\  {\bf 54} (2013) 082301
  [arXiv:1302.0378 [math-ph]].
%
\bibitem{Ablinger:2014bra}
  J.~Ablinger, J.~Bl\"umlein, C.G.~Raab and C.~Schneider,
  J.\ Math.\ Phys.\  {\bf 55} (2014) 112301
  [arXiv:1407.1822 [hep-th]].
%
\bibitem{AZ}
G.~Almkvist and D.~Zeilberger, 
J. Symb. Comp. {\bf 10} (1990) 571; 
\\
M.~Apagodu and D.~Zeilberger,
Adv. Appl. Math. (Special Regev Issue), {\bf 37} (2006) 139. 
%
\bibitem{Blumlein:2009cf}
  J.~Bl\"umlein, D.J.~Broadhurst and J.A.M.~Vermaseren,
  {\it The Multiple Zeta Value Data Mine},
  Comput.\ Phys.\ Commun.\  {\bf 181} (2010) 582
  [arXiv:0907.2557 [math-ph]].
%
\bibitem{GUESSHB}
M.~Kauers, {\it Guessing Handbook, Technical Report RISC 09-07, JKU Linz}.
%
\bibitem{Blumlein:2009tj}
  J.~Bl\"umlein, M.~Kauers, S.~Klein and C.~Schneider,
  Comput.\ Phys.\ Commun.\  {\bf 180} (2009) 2143
  [arXiv:0902.4091 [hep-ph]].
%
\bibitem{Blumlein:2017dxp}
  J.~Bl\"umlein and C.~Schneider,
  Phys.\ Lett.\ B {\bf 771} (2017) 31
  [arXiv:1701.04614 [hep-ph]].
%
\bibitem{Nogueira:1991ex}
  P.~Nogueira,
  J.\ Comput.\ Phys.\  {\bf 105} (1993) 279.
%
\bibitem{vanRitbergen:1998pn}
  T.~van Ritbergen, A.N.~Schellekens and J.A.M.~Vermaseren,
  Int.\ J.\ Mod.\ Phys.\ A {\bf 14} (1999) 41
  [hep-ph/9802376].
%
\bibitem{REDUZE}
  C.~Studerus,
  Comput.\ Phys.\ Commun.\  {\bf 181} (2010) 1293
  [arXiv:0912.2546 [physics.comp-ph]];\\
  A.~von Manteuffel and C.~Studerus,
  arXiv:1201.4330 [hep-ph].
%
\bibitem{Bauer:2000cp}
  C.W.~Bauer, A.~Frink and R.~Kreckel,   
  Symbolic Computation {\bf 33} (2002) 1,
  cs/0004015 [cs-sc].
%
\bibitem{FERMAT}
R.H.~Lewis, {\it Computer Algebra System {\tt Fermat}}, {\tt http://home.bway.net/lewis.}
%
\bibitem{Ablinger:2016yjz}
  J.~Ablinger, A.~Behring, J.~Bl\"umlein, A.~de Freitas and C.~Schneider,
  {\it Algorithms to solve coupled systems of differential equations in terms of power series},
  arXiv:1608.05376 [cs.SC];\\
  J. Ablinger, J. Bl\"umlein, A. de Freitas, C. Schneider , {\it A toolbox to solve coupled systems of differential 
and difference 
equations}, In: Proc. of 12th International
Symposium on Radiative Corrections (Radcor 2015) and LoopFest XIV (Radiative Corrections for the LHC and Future Colliders), 
PoS (RADCOR2015) 060, pp. 1-13 [arXiv:1601.01856].
%
\bibitem{Ablinger:2017tan}
  J.~Ablinger, A.~Behring, J.~Bl\"umlein, A.~De Freitas, A.~von Manteuffel and C.~Schneider,
  Nucl.\ Phys.\ B {\bf 922} (2017) 1
  [arXiv:1705.01508 [hep-ph]].
%
\bibitem{Vogt:2004mw}
  A.~Vogt, S.~Moch and J.A.M.~Vermaseren,
  Nucl.\ Phys.\ B {\bf 691} (2004) 129
  [hep-ph/0404111].
%
\bibitem{HSUM}
  J.A.M.~Vermaseren,
  Int.\ J.\ Mod.\ Phys.\ A {\bf 14} (1999) 2037
  [hep-ph/9806280];\\
  J.~Bl\"umlein and S.~Kurth,
  Phys.\ Rev.\  D {\bf 60} (1999) 014018
  [arXiv:hep-ph/9810241].
\end{thebibliography}
\end{document}